\documentclass[twocolumn,twoside]{revtex4}
\usepackage{graphicx}
\usepackage{fancyhdr}
\usepackage{epsfig}
\usepackage{amsmath}
\def\DAF{DA$\Phi$NE} 
\def\ifm#1{\relax\ifmmode#1\else$#1$\fi}
\def\ab{\ifm{\sim}}  
\newcommand{\eV}{{e\kern-.07em V}}

\newcommand{\MeV}{{\rm \,M\eV}}

\newcommand{\GeV}{{\rm \,G\eV}}

\newcommand{\ns}{{\rm \,ns}}

\newcommand{\ps}{{\rm \,ps}}
\newcommand{\mm}{{\rm \,mm}}
\newcommand{\cm}{{\rm \,cm}}
\newcommand{\mt}{{\rm \,m}}

\newcommand{\kl}          {\ensuremath{K_L}}
\newcommand{\ksl}         {\ensuremath{K_{S,L}}}
\newcommand{\kls}         {\ensuremath{K_{L,S}}}
\makeatletter
\newdimen\z@ \z@=0pt 
\newskip\z@skip \z@skip=0pt plus0pt minus0pt
\def\m@th{\mathsurround=\z@}
\def\ialign{\everycr{}\tabskip\z@skip\halign} 
\def\eqalign#1{\null\,\vcenter{\openup\jot\m@th
  \ialign{\strut\hfil$\displaystyle{##}$&$\displaystyle{{}##}$\hfil
    \crcr#1\crcr}}\,}
\makeatother
\newcommand{\BR}{B\kern -0.1em R}
\newcommand{\phif}{$ \phi$~{\em factory\/} }

\newcommand{\etal}{{\em et al.\/}}

\newcommand{\Vusfo}   {\ensuremath{|V_{\rm us} f_{\rm +}(0)|}}
\newcommand{\fo}   {\ensuremath{f_{\rm +}(0)}}
\newcommand{\Vus}   {\ensuremath{V_{\rm us}}}
\newcommand{\Vud}   {\ensuremath{V_{\rm ud}}}
\newcommand{\Vub}   {\ensuremath{V_{\rm ub}}}

\newcommand{\ele}[1]{\ensuremath{e^{{#1}}}}
\newcommand{\muo}[1]{\ensuremath{\mu^{{#1}}}}
\newcommand{\pai}[1]{\ensuremath{\pi^{{#1}}}}
\newcommand{\kao}[1]{\ensuremath{K^{{#1}}}}
\newcommand{\kneu}[1]{\ensuremath{K_{{#1}}}}

\newcommand{\kmudue}[1]{\ensuremath{K^{{#1}}_{\mu2}}}
\newcommand{\kpidue}[1]{\ensuremath{K^{{#1}}_{\pi2}}}
\newcommand{\ketre}[1] {\ensuremath{K^{{#1}}_{e3}}}
\newcommand{\kmutre}[1]{\ensuremath{K^{{#1}}_{\mu3}}}

\newcommand{\kltre}[1] {\ensuremath{K^{{#1}}_{{\rm l}3}}}
\newcommand{\Dkmudue}[1]{\ensuremath{K^{{#1}} \to \mu^{{#1}}\nu}}
\newcommand{\Dkpidue}[1]{\ensuremath{K^{{#1}} \to \pi^{{#1}}\pi^{0}}}
\newcommand{\Dketre}[1] {\ensuremath{K^{{#1}} \to \pi^0e^{{#1}}\nu}}
\newcommand{\Dkmutre}[1]{\ensuremath{K^{{#1}} \to \pi^0\mu^{{#1}}\nu}}

\newcommand{\bketre}[1] {\ensuremath{BR(K^{{#1}}_{e3})}}
\newcommand{\bkmutre}[1]{\ensuremath{BR(K^{{#1}}_{\mu3})}}
\newcommand{\lifet}[1]{\ensuremath{{\tau}_{#1}}}
\newcommand{\ff}[1]{\ensuremath{{\lambda}_{#1}}}
\newcommand{\rappo}[1]{\ensuremath{\Gamma(K^{{#1}}_{\mu3})/\Gamma(K^{{#1}}_{e3})}}
\newcommand{\rme}[1]{\ensuremath{R^{#1}_{\mu e}}}
\newcommand{\pb}     {\ensuremath{{\rm pb^{-1}}}}
\newcommand{\fb}  {\ensuremath{{\rm fb^{-1}}}}
\newcommand{\mdue}   {\ensuremath{m^{2}_{lept}}}
\newcommand{\mtrenta} {\ensuremath{p^{\ast}_{\mu}}}
\newcommand{\pstarp} {\ensuremath{p^{\ast}_{\pi}}}

\newcommand{\effi}[2]{\ensuremath{\varepsilon_{#1}^{#2}}}

\newcommand{\cdue}{\ensuremath{\chi^2}}
\begin{document}

\title{KLOE measurement of the charged kaon absolute semileptonic BR's}

\author{The KLOE Collaboration\\
F.~Ambrosino, A.~Antonelli, M.~Antonelli, F.~Archilli,
C.~Bacci, P.~Beltrame, G.~Bencivenni,
S.~Bertolucci, C.~Bini, C.~Bloise, S.~Bocchetta, V.~Bocci, F.~Bossi, P.~Branchini,
R.~Caloi, P.~Campana, G.~Capon, T.~Capussela, F.~Ceradini, S.~Chi, G.~Chiefari, P.~Ciambrone, 
E.~De~Lucia, A.~De~Santis, P.~De~Simone, G.~De~Zorzi, A.~Denig, A.~Di~Domenico, C.~Di~Donato,
S.~Di~Falco, B.~Di~Micco, A.~Doria, M.~Dreucci, G.~Felici,
A.~Ferrari, M.~L.~Ferrer, G.~Finocchiaro, S.~Fiore, C.~Forti, P.~Franzini, C.~Gatti,
P.~Gauzzi, S.~Giovannella, E.~Gorini, E.~Graziani, M.~Incagli, W.~Kluge, V.~Kulikov,
F.~Lacava, G.~Lanfranchi, J.~Lee-Franzini, D.~Leone,
M.~Martini, P.~Massarotti, W.~Mei, L.~Me\-o\-la, S.~Mi\-scet\-ti, M.~Moulson, S.~M\"uller, F.~Murtas,
M.~Napolitano, F.~Nguyen, 
M.~Palutan, E.~Pasqualucci, A.~Passeri, V.~Patera, F.~Perfetto, M.~Primavera,
P.~Santangelo, G.~Saracino, B.~Sciascia,
A.~Sciubba, F.~Scuri, I.~Sfiligoi, A.~Sibidanov, T.~Spadaro, 
M.~Testa, L.~Tortora, 
P.~Valente, B.~Valeriani, G.~Venanzoni, R.~Versaci,
G.~Xu}
\affiliation{Correspondig author: B.~Sciascia LNF - INFN, barbara.sciascia@lnf.infn.it}

\begin{abstract}
This paper is devoted to the measurement of the fully inclusive absolute branching ratios of the charged kaon semileptonic
decays,  \Dketre{\pm}$(\gamma)$  and \Dkmutre{\pm}$(\gamma)$.
The measurements have been done using a tag technique, employing the 
two-body decays \Dkmudue{\pm}\ and \Dkpidue{\pm}, and 
using a sample of about 410~\pb\ collected during the 2001 and 2002 data taking of the KLOE experiment
at \DAF\ the Frascati \phif.
The results obtained are \bketre{} = $0.04965~(38)_{Stat}~(37)_{Syst}$ and
\bkmutre{} = $0.03233~(29)_{Stat}~(26)_{Syst}$.

\end{abstract}

\maketitle

\thispagestyle{fancy}
The most precise verification of the unitarity of the CKM mixing
matrix is obtained today from the \Vus\ and \Vud\ values, neglecting
$|\Vub|^2$\ab0.00002.  With the KLOE detector we can measure all
experimental inputs to \Vus: branching ratios, lifetimes, and form
factors. 
Here we report about
the measurement of the fully inclusive absolute branching ratios of the charged kaon semileptonic
decays,  \Dketre{\pm}$(\gamma)$ (\ketre{\pm}) and \Dkmutre{\pm}$(\gamma)$ (\kmutre{\pm}),
using a sample of about 410~\pb\ collected during the 
2001 and 2002 data taking.

The first section~(\ref{sec:DaEKlo}) brefly describes the accelerator \DAF\ and the KLOE detector. 
The measurements have been done using a tag technique presented in section~\ref{sec:meto}.
The following two sections are dedicated to the selection of the tag samples~(\ref{sec:TagSele})
and the signal sample~(\ref{sec:SigSele}) respectively. 
MC efficiency has to be corrected using suitable Data and MC control samples, as described in section~\ref{sec:effi}.
The results are summarized in section~\ref{sec:result}, while the last section~(\ref{sec:KLOEvus}) contains
information about the \Vus\ extraction and the lepton flavour violation test, using all the \kltre{} KLOE results. 

\section{{\bf \DAF}\ and KLOE}
\label{sec:DaEKlo}
In the \DAF\ \ele{+}\ele{-} collider, beams collide at a center-of-mass 
energy $W\ab M(\phi)$.
 Since 2001, KLOE has collected an
integrated luminosity of \ab2.5 \fb.
Results presented below are based on 2001-02 data for \ab450 \pb.
The KLOE detector consists of a large cylindrical drift chamber surrounded by a 
lead/scin\-til\-la\-ting-fiber electromagnetic calorimeter. A superconducting coil around 
the calorimeter provides a 0.52~T field. The drift chamber~\cite{DC}
is 4~\mt\ in diameter and 3.3~\mt\ long. 
The momentum resolution is $\sigma(p_{T})/p_{T} \sim 0.4\%$. Two track vertices
are reconstructed with a spatial resolution of $\sim$ 3~\mm. The calorimeter~\cite{EMC}
composed of a barrel and two endcaps, covers 98\% of the solid angle.
Energy and time resolution are $\sigma(E)/E = 5.7\%/\sqrt{E(\GeV)}$ and
$\sigma(t) = 57~\ps/\sqrt{E(\GeV)} \oplus 100~\ps$.
The KLOE trigger~\cite{TRG} uses calorimeter and drift chamber information.
For the present analysis only the calorimeter signals are used. Two energy deposits
above threshold, $E>50$~\MeV\ for the barrel and $E>150$~\MeV\ for the endcaps, are required.

\section{Method of measurement}
\label{sec:meto}

The $\phi$ meson decays mainly into kaons: 49\% to \kao{+}\kao{-} and 34\% to \kneu{L}-\kneu{S} pairs.
The identification of the \kao{+} automatically tags the presence of a \kao{-}; the
same holds with reversed charges. In KLOE \kao{\pm} decays are tagged efficiently by the 
identification of a two body decay of one of the kaons. 
Both \Dkmudue{\pm} (\kmudue{\pm}) and \Dkpidue{\pm} (\kpidue{\pm}) decays
have been used for the measurements presented in this paper.

The use of a tagging technique gives the possibility to perform absolute branching ratio measurements.
The branching ratio of the signal decay can be extracted from:
$$ 
BR(Sig) = \frac{N_{Sig}}{N_{Tag}\effi{FV}{}\effi{Sig}{}}
\frac{\sum_{i}BR(i)\effi{Tag}{}(i)}{\effi{Tag}{}(Sig)}{\mbox,}
$$
where \effi{FV}{} is the geometrical acceptance 
of the fiducial volume, and \effi{SIG}{} is the signal reconstruction
and selection efficiency.
\effi{SIG}{} is measured directly using the MC, and 
corrected for the different
behaviors for Data and MC of quantities involved in the measurement like the tracking efficiency or
the single photon efficiency.

In principle, the capability of selecting a tag kaon does not depend on the decay mode of the
other kaon. In fact the geometrical superimposition of the ``tag'' and ``signal'' part of the \kao{+}\kao{-}
event, and the fact that the trigger and the background rejection and
tracking procedures look at the event globally, make the separation in two distinct
hemispheres arbitrary. In short, the \kao{\pm}\ tagging efficiency is not independent of the \kao{\mp}\ 
decay mode, and the tag bias has to be therefore carefully studied. 
The factor 
$$
\frac{\sum_{i}BR(i)\effi{Tag}{}(i)}{\effi{Tag}{}(Sig)}{\mbox,}
$$
should be equal to 1, if for each decay mode, $i$, the efficiency, $\effi{}{}(i) = \effi{}{}$.
This is referred to as the tag bias and can be measured only using MC. 
The correction due to the different behaviors for Data and MC of 
specific processes or sub-detectors has been measured in suitable dowscaled event samples.

\section{Tag selection}
\label{sec:TagSele}
The measurement of the branching ratios for the \kao{\pm} semileptonic decays
is performed using four data samples defined by different decay modes for the 
tagging kaon: \kmudue{+}, \kpidue{+}, \kmudue{-}, and \kpidue{-}.
This redundancy allows the systematic effects due to the tag selection
to be kept under control.

Tagging kaons are identified as tracks with momentum $70<p_K<130$ \MeV,
originating from the \ele{+}\ele{-} collision point~(I.P.). The kaon decay vertex must be
within a fiducial volume (FV) defined as 
a cylinder of radius $40<r<150$ \cm, and height |z|<130\cm, centered at the I.P.,
coaxial with the beams. The decay track, extrapolated to the
calorimeter, must point to an appropriate energy deposit.
\kmudue{\pm} (\kpidue{\pm}) decays are selected by applying a 3$\sigma$ cuts around 
the muon (pion) momentum calculated in the kaon rest frame, according to the
proper mass hypothesis.
For the \kpidue{\pm} tag, identification of the \pai{0} from the vertex is also required. Finally,
to reduce the dependency of the tag selection efficiency on the decay mode 
of the signal kaon, the energy deposits associated to the
tagging decay, are required to satisfy the calorimeter trigger.
In the analyzed data set about 60 million tag decays were identified and divided into the four tag samples.

The value of the tag bias, which accounts for the capability to separate each event 
into a ``tag'' and a ``signal'' part,
is shown in table~\ref{tab:tb}
separately per tag and per signal. 
This correction ranges from about -3\% to +4\% following the tag sample used. 
\begin{table*}[!htb]
  \begin{center}
  \begin{tabular}[c]{c|c|c|c|c}
                & \kmudue{+}     & \kpidue{+}     & \kmudue{-}     & \kpidue{-}     \\ \hline
    \ketre{}    & 0.9694(11)(49) & 1.0137(34)(52) & 0.9884(10)(47) & 1.0328(23)(32) \\ 
    \kmutre{}   & 0.9756(13)(50) & 1.0210(36)(52) & 0.9963(10)(48) & 1.0371(25)(32) \\ 
  \end{tabular}
  \caption{Tag bias per tag and per signal type, measured in MC and corrected for different Data-MC behavior
  Statistical and systematic errors are shown.}
  \end{center}
  \label{tab:tb}
\end{table*}

\section{Selection of semileptonic \kao{\pm} decays}
\label{sec:SigSele}
The reconstruction of a one-prong kaon decay vertex 
(performed with the same requests used in the identification of the tag)
is the first step of the selection 
followed by the identification of a \pai{0} associated to the decay vertex using a time of flight technique.
In doing this,
for each cluster not associated to a track, the kaon decay time, $t_i$, is calculated using the 
cluster time, $t_{clu}^i$, and the distance, $L_i$, between the vertex and the cluster position:
$t_i = t_{clu}^i - L_i / c$.  This time should be the same for two photons coming from a \pai{0} decay,
and at least a pair of clusters has to satisfy the condition $t_1-t_2~<~3\sigma_t$ 
($\sigma_t = \sigma_{t,1}\oplus\sigma_{t,2}$). 
Using the energy and the position of the clusters of the selected pair,
the $\gamma\gamma$ invariant mass is calculated and a 3$\sigma$
cut ($\sigma\sim$18\MeV) is applied around the nominal \pai{0} mass.
It is intended to remove events with accidental coincidence 
of clusters due to machine background clusters entering the calorimeter.

To isolate the \ketre{} and \kmutre{} decays, the lepton is identified by a time of flight 
technique. This requires that the lepton track is extrapolated to 
the calorimeter and geometrically associated to a calorimeter cluster.
Specifically, the kaon decay time estimated from the \pai{0} photons 
($t_{\pai{0}}^{decay}$) should be equal to the one estimated from the charged track ($t_{lept}^{decay}$), 
if the correct mass has been assigned to it.
The kaon decay time, estimated from \pai{0} photons, is: $t_{\pai{0}}^{decay} = t_i - L_i/c$.
For the lepton the proper track length and velocity are used: 
$t_{lept}^{decay} = t_{lept} - \frac{L_{lept}}{p_{lept}~c}~\sqrt{p_{lept}^2 + \mdue}$, where
$t_{lept}$ is the time of the cluster associated to the secondary track, and $p_{lept}$ and $L_{lept}$
are respectively the momentum, measured in the laboratory rest frame, and the length of this track.
The lepton mass is then obtained, imposing $t_{\pai{0}}^{decay} = t_{lept}^{decay}$:
$$
  \mdue ~=~ p_{lept}^2 \cdot 
  \left [ \frac{c^2}{L_{lept}^2}  \left (t_{lept} - t_{\pai{0}}^{decay}
    \right )^2 -1
  \right ] {\mbox .}
$$
To fight against the more abundant two body decays, the momentum of the secondary track is computed 
in the kaon rest frame, using the pion mass hypothesis. All events with \pstarp $>$195 \MeV\ are
rejected. Only poorly reconstructed \kmudue{} and \kpidue{} kinks, or \kpidue{} events with an 
early \pai{\pm} decay, survive this cut. 
These last events are rejected imposing to the lepton momentum (\mtrenta) calculated in the
center of mass of the \pai{\pm} (defined as missing momentum at the decay vertex, $P_K-p_{\pi^0}$), 
the condition \mtrenta $>$60~\MeV.
After all selection cuts, the contamination from non-\kltre{} events is less than $\sim$1.5\% in each tag sample. 
To obtain the number of signal events, a constrained likelihood fit is applied to the \mdue\ data spectrum 
using a linear combination of \ketre{} and \kmutre{} shapes, and of the background contribution.
MC shapes are corrected for Data/MC differences on the calorimeter timing. About 300\,000 \ketre{} 
and 160\,000 \kmutre{} have been found in the 2001-2002 data set.

\begin{figure}[!ht]
  \begin{center}
      \epsfig{file=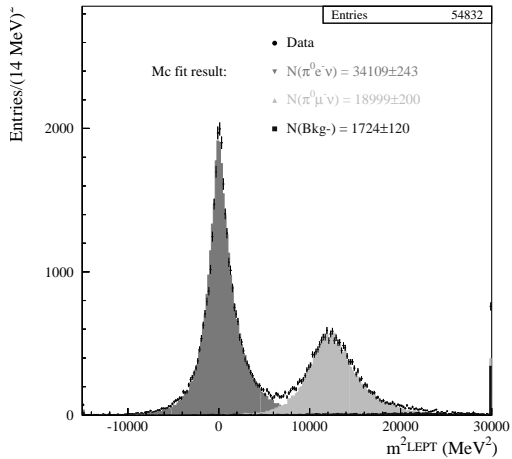,width=.9\linewidth}
      \caption{Fit results of \mdue\ distribution for the \kpidue{+} tag sample.}
    \label{fig:ke3peak}
  \end{center}
\end{figure}

In figure~\ref{fig:ke3peak}, the fit results of \mdue\ distribution for the \kpidue{+} tag sample is shown.
The \ketre{} signal component is evident as a narrow peak around zero, well separated from
the background. The \kmutre{} signal is the peak around the $m_{\mu}^2$ value.
Other tag samples show the same behavior.

\section{Efficiency evaluation}
\label{sec:effi}
The selection efficiency is measured on MC and corrected for relevant Data/MC differences, in particular the tracking efficiency
and the calorimeter clustering efficiency, for both photons and leptons.
For each correction, a control sample has been selected in such a way that the efficiency can be measured as function
of a suitable set of variables. 

For the tracking, a kinematic fit applied to a \kltre{} sample -statistically independent from the one used for the measurement-
has been used to extract the tracking efficiency as a function of the kaon polar angle, the decay vertex position,
and the lepton momentum.
The photon cluster efficiency has been measured as a function of the photon energy using a sample of \kpidue{\pm} decays.
A \kl$_{e3}$ event sample has been used to obtain the electron cluster efficiency as a function of the lepton momentum
and the incidence angle on the EMC. Finally the muon cluster efficiency has been extracted, as a function of the
same variables than the electron, using a combined sample of \kl$_{\mu3}$, \kpidue{\pm}
(with a well reconstructed \pai{}$\rightarrow$\muo{} kink), and \kmudue{\pm} events.

All the data/MC corrections are stable with respect to the variation of the cuts applied in selecting the control samples.

\section{Results}
\label{sec:result}
For all the relevant analysis aspects,
the systematic errors have been carefully evaluated for each tag sample and for each decay, taking correlations into account.
All contributions to the final error, both statistical and systematic,
are summarized in table~\ref{tab:l3errosummary} for both \ketre{} and \kmutre{} measurements.
\begin{table*}[t]
\begin{center}
  \begin{tabular}[c]{c|c|c|c|c|c|c|c|c}\hline
                    & \multicolumn{4}{c|}{\ketre{}} & \multicolumn{4}{c|}{\kmutre{}} \\ \hline
  Source            & \kmudue{+}& \kpidue{+} & \kmudue{-} & \kpidue{-}& \kmudue{+}& \kpidue{+} & \kmudue{-} & \kpidue{-} \\ \hline \hline
                    & \multicolumn{8}{|c|} {Statistical (\%)} \\ \hline
  Tag bias (TB)                     & 0.07    & 0.14    & 0.08     & 0.14   & 0.09   & 0.18   & 0.09   & 0.17  \\
  Cosmic veto corr. to TB           & 0.00    & 0.01    & 0.01     & 0.01   & 0.01   & 0.01   & 0.01   & 0.01  \\ 
  Machine Bkg corr. to TB           & 0.09    & 0.31    & 0.05     & 0.17   & 0.09   & 0.31   & 0.05   & 0.17  \\ \hline 
  Nuclear int. corr.                & 0.32    & 0.57    & -          & -        & 0.32   & 0.57   & -        & -       \\ \hline 
  Fit counting                      & 0.40    & 0.71    & 0.40     & 0.72   & 0.61   & 1.05   & 0.60   & 1.08  \\ \hline 
  \effi{}{} correction
                          & 1.17   & 1.67   & 1.24   & 1.78   & 1.61   & 2.25   & 1.12   & 2.27  \\ \hline\hline
                    & \multicolumn{8}{|c|} {Systematics: signal (\%)} \\ \hline
  \effi{}{} corr. (TRK)            & 0.54   & 0.54   & 0.53   & 0.53   & 0.44   & 0.43   & 0.43   & 0.43  \\
  \effi{}{} corr. (lepton clus.)   & 0.00   & 0.00   & 0.00   & 0.00   & 0.14   & 0.14   & 0.14   & 0.14  \\ 
  \effi{}{} corr. (\pai{0})        & 0.24   & 0.25   & 0.24   & 0.24   & 0.21   & 0.21   & 0.21   & 0.21  \\ 
  Fit                     & 0.13   & 0.19   & 0.35   & 0.15   & 0.03   & 0.16   & 0.19   & 0.06  \\ 
  Selection cuts          & 0.17   & 0.17   & 0.17   & 0.16   & 0.49   & 0.49   & 0.49   & 0.48  \\\hline \hline 
                    & \multicolumn{8}{|c|} {Systematics: acceptance (\%)} \\ \hline
  Nuclear int. corr.      & 0.18   & 0.39   &  -       &    -     & 0.18   & 0.39   &  -       &  -      \\
  \lifet{\pm}             & 0.09   & 0.09   & 0.09   &  0.09  & 0.09   & 0.09   & 0.09   & 0.09   \\\hline \hline  
                    & \multicolumn{8}{|c|} {Systematics: tag bias corrections (\%)} \\ \hline
  Machine Bkg             & 0.36   & 0.06   & 0.37   & 0.05   & 0.36   & 0.06   & 0.37   &  0.05   \\
  Cosmic veto             & 0.04   & 0.02   & 0.03   & 0.04   & 0.04   & 0.02   & 0.03   &  0.04   \\
  Nuclear int.            & 0.09   & 0.13   &  -       &    -     & 0.09   & 0.13   &  -       &   -      \\ \hline  \hline 
                & \multicolumn{8}{|c|} {\bf Total (\%)}   \\ \hline
                & {\bf 1.49 } & {\bf 2.08 } & {\bf 1.54 } & {\bf 2.03 }& {\bf 1.95 } & {\bf 2.71 } & {\bf 1.52 } & {\bf 2.63 } \\ \hline
  \end{tabular}
  \caption{Summary of all fractional contributions to the error on \ketre{} and \kmutre{} branching ratio measurements.}\vglue2mm
  \end{center}
  \label{tab:l3errosummary}
\end{table*}
The final errors are dominated by the statistical error of the efficiency corrections: 
the tracking is the dominant contribution for the \ketre{} measurement, while for \kmutre{},
tracking and muon cluster corrections are at the same level.
Final fractional accuracy ranges, depending on the tag sample, from 1.5\% to 2.1\% for the \ketre{}, 
and from 1.5\% to 2.7\% for the \kmutre{} measurements.

The \cdue\ for the 4 independent-tag measurements is 1.62/3, with a probability of about 65\% for the \ketre{},
and 1.07/3, with a probability of about 78\% for the \kmutre{} decays.
The averages of the four results are:
\begin{align*}
\BR(\ketre{-}) & = \left(4.946 \pm 0.053_{Stat} \pm 0.038_{Syst}\right)\times10^{-2} \\
\BR(\ketre{+}) & = \left(4.985 \pm 0.054_{Stat} \pm 0.037_{Syst}\right)\times10^{-2} \\
\BR(\ketre{ }) & = \left(4.965 \pm 0.038_{Stat} \pm 0.037_{Syst}\right)\times10^{-2} {\mbox ,}
\end{align*}
for the \ketre{}, and 
\begin{align*}
\BR(\kmutre{-}) & = \left(3.219 \pm 0.047_{Stat} \pm 0.027_{Syst}\right)\times10^{-2}\\
\BR(\kmutre{+}) & = \left(3.241 \pm 0.037_{Stat} \pm 0.026_{Syst}\right)\times10^{-2}\\
\BR(\kmutre{ }) & = \left(3.233 \pm 0.029_{Stat} \pm 0.026_{Syst}\right)\times10^{-2} {\mbox ,}
\end{align*}
for the \kmutre{}. The \cdue\ between different charge measurements is 0.17/1, with a probability of about 68\% for the \ketre{},
and 0.12/1, with a probability of about 73\% for the \kmutre{} decays.
Final \BR\ results have a fractional accuracy of 1.1\% for the \ketre{} and of 1.2\% for the \kmutre{} decays, and
are in agreement within the errors with the KLOE preliminary results~\cite{Kpml3_KLOE05}.

The \lifet{K} value affects the \BR\ values via the geometrical acceptance evaluation. From suitable MC sample
the \BR\ dependency on the \lifet{K} has been estimated: 
\BR(\kltre{},\lifet{}) = \BR(\kltre{},\lifet{}$^0$)$\cdot$(1-0.45(\lifet{K}-\lifet{}$^0$)/\lifet{}$^0$), with
\lifet{}$^0$ = 12.36\ns,\BR(\ketre{},\lifet{}$^0$) = 4.968\%, and \BR(\kmutre{},\lifet{}$^0$) = 3.234\%.
The final results are evaluated using the PDG06~\cite{PDG06} fit value \lifet{} = 12.384(24)\ns.

While the correlation between the \ketre{} and \kmutre{} signals induced by the fit procedure is low (about 1\%),
a correlation is caused by the corrections to the tag bias, which are equal for the two signals, by the same
Data/MC corrections for the tracking and the clustering, and finally by the selection cuts.
Excluding the error coming from \lifet{K} value, the complete information is contained in the error matrix:
\[ \left( \begin{array}{cc}
0.2780 & 0.1268 \\ 
0.1268 & 0.1510
\end{array} \right) \times 10^{-6} {\mbox ,}\]
from which a 62.74\% of correlation between \ketre{} and \kmutre{} can be extracted.

For \kmutre{} only also the dependency on \ff{0} has to be considered. The final \bkmutre{} has been evaluated
using \ff{0}$\sim$0.015 and the relation \bkmutre{} = 0.0327 - 0.0230~\ff{0}, obtained from a dedicated MC study.
The limited knowledge of \ff{0} value gives a negligible contribution to the systematic error.

The \BR\ are completely inclusive. The radiated photon acceptance is determined with a generator that uses
the soft-photon approximation to sum the amplitudes for real and virtual processes to all order of $\alpha$~\cite{MCgene_Gatti}.

The ratio \rme{}=\rappo{} has been measured in the four tag samples used for the \BR\ measurements.
With respect to the \BR, \rme{} benefits of smaller tag bias corrections ranging from 0.4\% to 0.8\%
depending on the tag sample.
The correlation between \ketre{} and \kmutre{} coming from the fit procedure and from the efficiency
corrections has been taken into account in calculating the error on \rme{}. As in the \BR\ determination,
the errors are dominated by the statistics of the efficiency corrections. The average of the four tag sample measurements
is \rme{KLOE}$=0.6511~(46)_{Stat}~(73)_{Syst}$.
This measurement has a fractional error of about 1.3\% and is in agreement within the errors with the theoretical
prediction: \rme{SM}$=(1-\delta_K^\mu)/(1-\delta_K^e)\cdot(I_K^\mu)/(I_K^e)=0.6646(61)$,
where the integrals ($I_K^e$ and $I_K^\mu$) and $\delta_{SU(2)}$, $\delta_{em}$ corrections are from~\cite{FlaviaNet06}.

\section{\Vus\ determination from KLOE results}
\label{sec:KLOEvus}
\begin{figure}[!ht]
  \begin{center}
      \epsfig{file=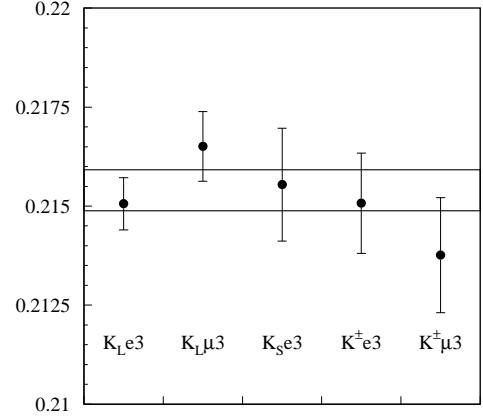,width=.8\linewidth}
      \caption{\Vusfo\ measurements from KLOE \kltre{} results.}
    \label{vusfzero}
  \end{center}
\end{figure}

The BR's of the semileptonic \kao{\pm} decays, together with the already published results 
for the semileptonic decays of the \kneu{L}~\cite{BR_KL} and \kneu{S}~\cite{BR_KS},
allow five independent determinations of the observable \Vusfo, as shown in Fig.~\ref{vusfzero}.
All inputs, but \kltre{} branching ratios, needed to extract \Vusfo\ are from~\cite{FlaviaNet06}.
The average of the five \Vusfo\ values, taking correlations into account,
gives \Vusfo\ = 0.2154(5), with \cdue\ = 4.37/4, and has a probability of $\sim$36\%.
Using \fo = 0.961(8) from~\cite{Leut}, \Vus\ is 0.2241(19). This is compatibile with the Unitarity
at 1.5~$\sigma$, in fact using \Vud=0.97377(27)~\cite{vudvalue}, \Vud$^2$+\Vus$^2$-1 is equal to -0.0015(10).
\Vusfo\ can be evaluated also by charge state. Using \ksl\ values we obtain 0.2155(6), while using
\kao{\pm} we obtain 0.2146(12). These two determination have a \cdue\ of 0.48/1 with a probability of 49\%.

The comparison of \Vusfo\ results by charge state, allow to evaluate empirical $\Delta$SU(2) correction.
This is 1.88(58)\%, and has to be compared with the $\chi_{PT}$ prediction of 2.31(22)\%. The two evaluations are
in agreement within the errors.

Lepton universality can be tested comparing the \rme\ value from the \kltre{} measurements, \rme{Obs},
and the one from the SM expectations, \rme{SM}, defined in the previuos section.
Defining $r_{\mu e}$ = \rme{Obs}/\rme{SM}, and using the \BR\ KLOE results we obtain for neutral kaon 
$r_{\mu e}$(\kls) = 1.013(9), and for charged kaon $r_{\mu e}$(\kao{\pm}) = 0.988(11).
Both results are compatibile with 1 within the errors,
and average to $r_{\mu e}$=1.003(7), with a \cdue\ of 3.60/1 and a probability of 5.8\%.

\end{document}